# A Philosophical Understanding of Representation for Neuroscience

Ben Baker[1], Benjamin Lansdell[2], Konrad Kording[3]

**Abstract:**
Neuroscientists often describe neural activity as a representation of something, or claim to have found evidence for a neural representation. But what do these statements mean? The reasons to call some neural activity a representation and the assumptions that come with this term are not generally made clear from its common uses in neuroscience. Representation is a central concept in philosophy of mind, with a rich history going back to the ancient period. In order to clarify its usage in neuroscience, here we advance a link between the connotations of this term across these disciplines. We draw on a broad range of discourse in philosophy to distinguish three key aspects of representation: correspondence, functional role, and teleology. We argue that each of these aspects are implied by the explanatory role the term plays in neuroscience. However, evidence related to all three aspects is rarely presented or discussed in the course of individual studies that aim to identify representations. Overlooking the significance of all three aspects hinders communication in neuroscience, as it obscures the limitations of experimental paradigms and conceals gaps in our understanding of the phenomena of primary interest. Working from this three-part view, we discuss how to move toward clearer communication about representations in the brain.

## 0 Introduction

Experimental data in neuroscience are interpreted and communicated in terms of certain key concepts, such as those of coding, representation, and computation. The question of how these concepts apply and relate to one another in different cases is one that requires taking a philosophical perspective on the practice of neuroscience. This paper works in that vein, and is focused on the idea of representations in neural systems.

Neuroscientific publications frequently call a pattern of neural activity a representation of something, however there is no widely agreed-on definition of "representation," and the implied meaning of the term across different sub-fields and individual scientists is somewhat vague and inconsistent. A survey of authors within two areas of research in neuroscience revealed significant variation in respondents' ways of describing what is implied by "representation" (Vilarroya 2017). Recent theoretical work has pointed out a disconnect between the kind of understanding suggested by claims involving neural representation and the kind of observations typically offered as evidence of representations (Brette 2018, Kragel et al. 2018, Jones & Kording 2019). Relatedly, a forceful criticism has also been made about the lack of clarity around representations in Cognitive Science and AI (Harvey 2008), as well as a critique of common ways of moving between quantitative expressions and qualitative ones in psychology (Yarkoni 2020). This all provides a rationale for neuroscientists to step back from any particular data or models and carefully assess what representations in the brain entail.

[1] Department of Neuroscience, University of Pennsylvania
[2] Department of Bioengineering, University of Pennsylvania
[3] Department of Neuroscience, Department of Bioengineering, University of Pennsylvania, CIFAR

The idea of representation as important to the understanding of intelligent agents is a focal topic in contemporary philosophy and also, in some form, going back to the ancient period. While philosophers have offered varied and competing accounts of representation, a few broad themes can be identified in the theoretical challenges that philosophers have posed and tried to answer. These discourses directly concern the framing assumptions that are obscure or imprecise in how the term is used in neuroscience. Whether it is because philosophers have not made their ideas sufficiently accessible to neuroscientists, or because neuroscientists have not sufficiently attended to the philosophical issues raised by their work, there is considerable room for neuroscience to be better informed about how philosophy has treated the issue of representation (Laplane 2019). This paper draws broadly on philosophical literature while closely examining the common methods and tools of neuroscience in an effort to disambiguate the notion of representation in a way that will best serve the communicative interests of neuroscientists.

We do this by distinguishing three aspects of representation - three conditions implied in the identification of representations if they are to play the key explanatory role suggested by their central place in common claims about what some part of the brain does. Briefly, the first aspect is a relation of correspondence (or information-carrying) between a representation and the thing it represents, the second relates a representation causally to behavior via a functional role, and the third puts representation in a teleological (purposive) context. We discuss how these three ideas are highlighted in the work of various philosophers, and we show how each is involved in the kind of explanation that representations are asked to provide in neuroscience. We note how each of these aspects is plausibly at least implicit in many neuroscientists' use of the term "representation," but also how the field has tended to disregard the significance of our second and third aspects. We discuss how embracing the full, three-aspect notion of representation and being attentive to the implications of each can allow for clearer expression of hypotheses and findings in neuroscience. In sum, we try to develop a more refined understanding of this core concept, enabling neuroscientists to direct and communicate their research more effectively.

In section 1, we discuss the range of meanings that "representation" has been most closely linked to in neuroscience, and we suggest how the notion of representation we outline will serve general neuroscientific aims. In section 2 we discuss each aspect in turn; we put each in the context of philosophical views that define or emphasize it, relating each aspect to examples and clarifying the significance for neuroscience. Lastly, there is a brief section of "Objections and Replies."

# 1   Common Uses of "Representation" in Neuroscience

Neuroscience as a whole uses a varied and somewhat informal notion of representation. On first pass, a neuroscientist might define a representation as a state or set of states within the brain that serves as a stand-in for some relevant feature of the world, acting as a kind of description or token for something (deCharms and Zador 2000). This intuitive definition can be broken into two parts; that the representing states within the brain in some way relate to the things being represented, and that these states are used by the animal -- they interact with other brain processes in a way that relates to the animal's behavior. Thus, identifying a neural representation of some object $F$ is supposed to help us understand how the animal whose brain it is does interesting, $F$-related things. Starting from this loose definition, in this section we distinguish some different contexts in which neuroscientists commonly use the term "representation," and discuss the ideas most closely associated with its use in these contexts. Doing so conveys a sense of

"representation" that is somewhat shallow or incongruous. In this way, we flesh out what neuroscientists aim to do with the idea of representation and illuminate the broad issue we take on in this paper.

There are at least three distinct contexts in which neuroscientists commonly use the concept of representation. The first is based simply on the examination of an animal's behavior. The informal definition above suggests that an animal uses representations in the production of adaptive behavior, supporting the idea that an animal's behavior can tell us something about things represented in parts of the animal's brain. More specifically, behavior that is not an immediate response to stimulation seems indicative of some form of representation, since representations are presumed to mediate more flexible and temporally extensive behaviors. For example, if an animal possesses a representation of a predator outside of its visual (or auditory, or olfactory...) field, it can use this representation to make predictions about the predator's future location and act accordingly, even without immediate sensory evidence about the predator's location. Behaviors that have an anticipatory component in this sense naturally lead to hypotheses about representations, and indeed, this has been proposed as a basis for defining representations (Brette 2018; Mirski and Bickhard 2019). Not all animals possess this sort of anticipatory behavior: animals whose behavior is largely reactionary do not invite the same inference about internal representations driving the animal's clever response to distant parts of the world outside it. The Hydra, for example, appears to have a relatively reactionary behavioral repertoire (Han et al. 2018). Although observations of behavior cannot, by themselves, answer how an animal's nervous system implements the relevant representations, behavior is a key component driving some hypotheses about what things animals represent (Krakauer et al. 2017).

A second common context for thinking about representations is in observing a link between the properties of neurons (or populations thereof) and salient features of an animal's environment. This perspective is one that focuses first on measured relationships between neural activity and the environment, and uses the idea of the relationship between a representation and a thing represented to more fully characterize the neural-environmental connection. Neuroscience experiments often involve recording the activity of neurons and measuring how that activity correlates with changes in stimuli or behavior that experimenters can observe or control. In essence, such studies identify patterns of covariation between neural responses and whatever experimental variable is being set, either stimulus parameters or quantified behavior. The analysis of this usually involves tuning curves (deCharms and Zador 2000, Kriegeskorte and Wei 2021). In contrast to the above discussion on behavior, it is easiest to quantify correlations to things in the world that an animal is currently experiencing. Thus, this second approach to neural representation is most associated with work on early sensory areas, where activity largely only depends on incoming stimuli, providing the cleanest connection between neural activity and features in the world (Purves et al. 2018). Neuroscientists studying vision, for instance, may record from neurons in V1 and observe how their activity is correlated with visual stimuli it was concurrently presented with. It is common to infer representational content based on the strength of these correlations, although the soundness of such inferences has been challenged within neuroscience (Brette 2018). That is, the neural activity may simply be taken to be a representation of whatever environmental variable the activity is well correlated with.

Many neuroscientists also go further and interpret such correlational experiments in terms of a communication metaphor, which can be quantified with Shannon's information theory (Simoncelli and Olshausen 2001). This approach interprets neural activity as transmitting one among multiple possible signals that are sent through presynaptic connections and received by postsynaptic neural structures. This interpretation allows for asking how much information about a given external variable is contained in the activity. Often neurally transmitted information is further interpreted to be an 'encoded' version of features of the outside world. This notion of a 'neural code' assumes that the relevant information from the environment is transduced by peripheral sensory mechanisms and encoded into the format in which the neurons communicate, and assumes further that this neural activity is subsequently decoded by downstream processes in the brain. The notion of a neural code is ubiquitous in neuroscience, and may often be used interchangeably with that of neural representation (e.g., Allen et al. 2017; Hirokawa et al. 2019; Ruff and Cohen 2019; Kaplan & Zimmer 2020). However, while the neural code concept allows for a lot of beautiful theory to do with channel capacity and efficiency that can drive hypotheses about how neurons send information (Rieke 1997), it generally does so by focusing almost exclusively on how neural activity correlates with external stimulus features, and not on a neural population's role in driving adaptive behavior. Thus, this approach seems to ignore part of the rough definition we offered above, and seems to stand in tension with the way some neuroscientists think of representation as, first and foremost, involved in flexible or anticipatory behavior.

A third context in which neuroscientists often use the idea of representation is in relation to variables that describe properties internal to the information-processing done by the nervous system. Many systems neuroscientists are especially interested in steps of information processing that support decision making, memory, reasoning, etc. Following Marr (1982), this leads to thinking about what algorithms the brain may be using to achieve some computation, and how such algorithms may be implemented. A computation, in this sense, is some task the organism performs or some problem it solves, which it is able to do by following the steps that make up an algorithm, where different algorithms might achieve the same computation.[4] The steps of that algorithm are presumed to be implemented by various neural activities, and neuroscientists working in this framework look for evidence that some parts of the brain implement a particular algorithm. Such algorithms commonly involve representations. The contents of these representations (and interactions among them) allows one to analyze the larger computation into parts that make it up. Often these contents do not pertain to external environmental variables at all. Rather, neural activity might represent something about the state of the system that carries out the algorithm. As an example, consider the ramping of spiking activity in the lateral intraparietal cortex (LIP) when primates are presented with multiple-moving-dot stimuli whose average motion is ambiguous between directions. According to one theory, the spiking rate of neurons in LIP represent the integrated evidence for the stimuli moving in one direction or the other (Shadlen and Newsome 2001; Pillow et al. 2008; Xuelong Zhao and Kording 2018; Latimer et al. 2015)). So these neurons are taken to represent something about the computational system itself - how much evidence it has accrued - rather than representing something about an external object. Studying representations in this third sense thus requires the additional theoretical framework associated with computational algorithms and their neural implementation, and is not clearly centered on a relationship between an animal and its environment, the way the previous approaches were.

In sum, we have just illustrated three prominent senses in which neuroscience uses the concept of representation -- one based on behavior, one based on relations between neural activity and the world, and one used in the exploration of computational algorithms in the brain. Each approach most naturally focuses on different types of representations. In studying representation from behavior, one focuses on anticipatory action and capacities needed to store information about things no longer being presented to an animal; in studying representation from correlations with neural activity, one focuses on immediate stimuli or motor behaviors and early sensory and primary motor areas; in studying representations involved in computation, one focuses on specific algorithms and often on representations of internal variables. Each approach suggests forms of evidence for representations, and each has different motivating assumptions. A behavior-based approach may make no commitments about how a given representation is implemented. An approach based on neural correlates may come with unstated assumptions about a role for such correlates in shaping behavior, but may not directly examine that role. A computation-based approach revolves around the issue of how algorithms are implemented in a nervous system. In this paper we leave aside the topic of neural computation - we are concerned with neural representations of things in the outside world, irrespective of how they might figure in computations. The key point here is that a neuroscientific study generally will not make explicit which sense of representation is being used, leaving it unclear exactly what sort of evidence is supposed to warrant the use of the word representation. The result of these heterogeneous and unarticulated assumptions is that an imprecise and vague notion of representation pervades neuroscientific communications.

---

[4] Some philosophers and cognitive scientists of note have employed a different notion of computation, which is defined just as the rule-governed manipulation of symbolic states, irrespective of any semantic contents that might be associated with those symbols (Newell et al. 1958, Fodor 1975, Haugeland 1978**,** Chalmers 1994**,** Piccinini2008**)**.

Representation is also foundational in much contemporary work in philosophy of mind, and some of the thinking in neuroscience is plausibly already influenced by philosophical conceptions of representation. (The movement of central ideas across disciplines is a difficult question we will not speak to directly here). We believe neuroscience would benefit from being more informed on philosophical approaches to internal representations in thinking beings, and we try to help bridge this gap.

## 2      Three Aspects of Representation for Neuroscience

Now that we have an outline of some common ways that representation is used in neuroscience, we turn to take an overview of the idea as it has been analyzed by philosophers. The scope and diversity of philosophical literature on representation is far too great for us to succinctly summarize here, but we have tried to capture those parts of the literature that make clear connection to the physiology or functional description of animals or artificial systems. We draw on a wide range of views pertaining to representations in nervous systems that have generated significant discourse in recent decades, and we also point out roots or analogues of these perspectives in much older philosophical work. We then relate these philosophical ideas back to neuroscience by putting them in the context of concrete examples. In so doing, we will distinguish three main aspects of representation that we argue are implicated in the explanatory role they are supposed to play in neuroscience. We illustrate that neuroscientific use of the term "representation" often seems concerned with all three aspects at some level, although often the latter aspects are somewhat disregarded. We suggest that neuroscience can gain clarity around this issue by adopting a notion of representation that more fully appreciates the significance of all three, and accordingly being careful to think and write about how all three are entailed in the identification of representations in a nervous system.

In broad terms, philosophers commonly understand representations to be entities that have semantic content, which can come in various types or modes. In other words, a representation is *about* something, and it represents what it is about in one of various ways. For example, a map, a painting, and an utterance could all represent the same place in different ways. Representationalism is the idea that mental processes essentially involve some form of representation, and this idea has been prevalent in philosophy for a long time. In the era of brain research there have been various philosophical accounts of what representation involves, which generate different assessments of the way neuroscience experiments try to reveal representation (Field 1978; Fodor 1987; Papineau 1987; Dretske 1988; Hatfield 1991; Churchland and Sejnowski 1990; Markman and Dietrich 2000; Millikan 2001; Grush 2004; Ryder 2004; Eliasmith 2005; Rupert 2018; Shea 2018). In the previous section, we identified significant variation in neuroscientists' uses of the idea of representation. Others have raised philosophical challenges regarding the way representations are broadly understood in neuroscience and cognitive science (Van Gelder 1995; Ramsey 2007; Harvey 2008; Egan 2014), and there have also been recent critiques from within neuroscience identifying a theoretical gap between what typical experiments reveal about neural activity and the explanatory role neural activity is interpreted to play (Kragel et al. 2018, Jones and Kording 2019; Brette 2018). We build on these accounts and critiques to distinguish three aspects of the role representations play in neuroscience.

We want to note that our view does not prescribe that all neuroscientists should be equally interested in investigating all three aspects at a time or in general. Different parts of the field are in position to further our knowledge about different aspects in different degrees. Our proposal that representation involves all three is meant to reveal how, within the singular notion of representation, we capture different parts of what it is about neural systems that is relevant to our understanding of the creatures that have them. Any subset of these aspects can be well described in terms that do not involve representation - roughly, in terms of information, functional role, and purpose. This does not mean that neuroscientists should be forbidden from using the term "representation" unless there is complete certainty about all three aspects. For example, work that only directly probes one aspect could explicitly state the assumptions that are involved in identifying some neural activity as a representation, making it clear how ascribing the status of representation depends on more than the experimental data acquired. As we illustrate, neuroscience already does this to an extent by citing other research or reasonable conjecture that points to the aspects not revealed by the study at hand. However, as we already argued, there is considerable lack of unanimity and consistency in the implicatures suggested by the term "representation." Our claim is that the three aspects below, which are all well-grounded in the history of philosophical thinking about representation, are each essential to the way neuroscience can most clearly use representations to reveal the workings of the brain.

**Aspect 1: Correspondence**

Our first aspect describes representations as matching or corresponding to the things they represent. So one says, for instance, it is partly by having inner states that match or correspond to red fruit in an animal's environment that the animal can recognize such fruit. More specifically, one might say there must be an isomorphism between internal representations and the external states they represent, which would result in an observable correlation in the way internal and external states vary. In general, this aspect says that if some neural activity $N$ represents some event or feature of the world $F$, then one should be able to find evidence of a correspondence between $N$ and $F$ in the form of correlational data.[5]

---

[5] Here we mean "correlation data" in a quite general sense, encompassing any form of statistical dependence one might investigate.

This kind of correspondence is central to representations as they figure in typical neuroscientific research. A classic example is given by Hubel and Weisel's (1959) original finding of receptive fields associated with cells in the visual cortex of a cat. In short, these cells' activity were shown to correspond to stimuli in specific portions of the cat's visual field via experiments that revealed correlations between recorded cell responses and stimulus locations. This line of work revealed in detail how features of the animal's visual environment are reflected in the organized activity of certain parts of the brain, making a huge contribution to our understanding of the neural processes that underlie visual perception and behavior more broadly.

This notion of correspondence is also deeply rooted in the history of philosophical thinking about representations. A prevailing view among European philosophers in the early modern period was a resemblance-based view of representation. For example, Locke and Hume conceived of internal "impressions" or "ideas" of external objects as representing those objects in virtue of resembling them (Locke 1690/1948; Hume 1739/1978). Isomorphism is important to philosophers of mind at least as far back as Aristotle's *De Anima*, where Aristotle conceives of registration by perceptual organs on analogy to the symbol that remains in the shape of some wax after a signet ring has been pressed into it (Book II, Chapter 12). Philosophers found compelling the notion that clever behavior depended on internal states whose "shapes" and interactions mirrored things in the outside world long before any were aware of the brain's capacities to implement complex formal relationships.

There have been some that try to define representation just in terms of a kind of correspondence, such as a measure of mutual information (Usher 2001) or higher-order structural similarities (O'Brien and Opie 2004). However the bulk of the philosophical literature on the topic suggests that correspondence alone does not establish the explanatory relevance that representations are supposed to have. As we will discuss in connection with the remaining two aspects, most accounts appeal to theoretical principles other than, and often in addition to correspondence to determine what counts as a representation. The first aspect is insufficient for representation because, in short, any statistical or morphological form of correspondence is sure to exist between neural processes and all sorts of things they are not representations of. For instance, circadian rhythms can be found in neurons and other cells throughout the brain, but this does not make these cells representations of the time of day. One can demand a stronger correspondence by some measure, but by most accounts no such measure will remove the distance between correspondence and representation and the relation of "aboutness" that it entails. This is brought out most clearly by cases where the behavioral stakes are high and the cost of a false negative is significantly different from that of a false positive. Consider, for example, that many animals respond with avoidance behaviors to relatively simple features of visual stimuli, such as a looming dark object (Schiff, Caviness, and Gibson 1962; Xinyu Zhao, Liu, and Cang 2014). Observable neural structures over short time-spans may reliably lead to avoidance behavior, as in the case where these neural activities correspond with "looming" described in strictly visual terms. However it is widely assumed in such cases that there is represented content that merits the avoidance behavior - a representation of something that poses a threat, which looming patterns of light do not, strictly speaking. In such cases, our confidence in the existence of a representation evidently comes from something other than the strength of a correlation, since the correlation strength is low between the neural activity and what it is meaningfully related to in the context of the animal's life. By many philosophical accounts, correspondence (often in information-theoretic terms) is a crucial piece of the puzzle of representation, but some further theoretical resources are needed to determine which correspondences are relevant (Dretske 1981; Fodor and Fodor 1987; Bickhard 2000; Millikan 2001; Shea 2018; Baker 2021).

The notion that brain activity must in some way reflect the things being represented is well explored in neuroscience and might be called a predominant focus. Most systems neuroscience experiments correlate neural activity with features of an animal's environment, e.g. stimuli or movement, and theorize on the basis of such models. At least since the work of Edgar Adrian, the observed correspondence between neural activity and stimulus features was recognized as important, and as potentially related to representation or coding. In fact, the most recent Nobel prize on the finding of correlations of neural activity with external variables just happened in 2014 (Burgess 2014) while the first one was awarded in 1963 (Huxley 1963). Experimentally, such observations are possible even with single electrode recordings in conjunction with presented stimuli or observed behavior. Thus, they have formed the basis for many theories and conceptions about how the brain represents features of the world and how the brain generates adaptive behavior.

**Aspect 2: Functional Role**

Our second aspect follows directly from the sort of explanation that representations are supposed to provide for; representations help us to make sense of how an animal acts on the basis of information it has access to that is relevant to its needs and wants. Representations of features of the world that pose opportunities and threats are part of what drive creatures to act in regular ways - toward what they need, away from what is dangerous, etc. Thus, representations typically figure in causal explanations of behavior. (DeCharms and Zador, 2000; Gallistel and King, 2009). For example, a representation of a predator might be offered as part of a causal explanation of an animal's flight. Generally, this aspect says that, to support a claim that neural activity $N$ is a representation of $F$, one should have reason to believe that $N$'s correspondence with $F$ could be causally involved in some $F$-related behavior(s).

Consider, for example, one study that offers a detailed hypothesis about representations of the direction of a higher concentration of nutrients in neural activity in the worm, *C elegans*, (Soh et al., 2018). Simplifying somewhat, the authors present evidence to support a model on which the difference between the firing rates of two pairs of interneurons serves to represent the gradient of NaCl across the area in front of the creature's nose-tip. The information about this chemical gradient is, of course, present in other neural activity besides that of the two interneurons; some sensory neural activity upstream from the proposed representation transmits the information to the interneurons, which then affect the activity of other neurons. So the NaCl gradient will correlate with patterns of activity besides that of the two interneurons. In other words, the relevant information is, in principle, decodable from various structures other than the hypothesized representation by the activity of two neurons in particular.

We find it not exactly clear from the Soh et al.'s use of the term what underlies their application of "representation" to the interneuronal difference in activity (this point of vaguess we find common, which partly motives our paper). The introduction of the term by the authors says nothing to explicitly suggest that anything further than a statistical correspondence (Aspect 1) is needed to qualify the interneuronal difference in activity as a representation. However, their hypothesis about the content of this representation hinges on observations as to how motor output appropriate to following the NaCl gradient depends on the activity of these interneurons. So the control of locomotion is key to these authors' approach to identifying the representations in this system. Therefore one might think that part of Soh et al.'s reason to identify the interneuron activity as a representation of the NaCl gradient comes from the role that activity apparently has on behavior involving the NaCl gradient (Aspect 2).

Let us briefly consider a second example, with the promise to give a somewhat fuller picture of each in proceeding sections. In a classic work on "place cells," Burgess and O'Keefe (1996) examine, in the rat hippocampus, "place cells that code for the directions of goals during navigation" (abstract). In short, when rats learn a maze through exploration involving the discovery of rewards, part of the hippocampus will become organized in a way so that its activity corresponds to the rat's location in the maze (Aspect 1). Burgess and O'Keefe offer a model of the role that networks in the rat hippocampus might play in navigation. In their model, the rates of place cells firing represent spatial relations between the rat and some goal location(s) in its environment. The authors are clear that they are interested in neural activity that plays a role in shaping behavior (Aspect 2). What they offer in the first place is a model of how a rat can (learn to) navigate a maze, so representations of "place" are formulated in the context of explaining reward-oriented navigation. Also, the authors appeal to impairments in navigation caused by hippocampal lesions to support their model. Thus, on one natural reading of Burgess and O'Keefe's work, ascribing the content about "place" to these hippocampal cells implies they have some observable effect on a behavior involving "place." Going back further in place-cell research, to O'Keefe and Dostrovsky's (1971) first results, the authors took their findings to "suggest that the hippocampus provides the rest of the brain with a spatial reference map." So in this prior work, even though no data about navigation behavior was recorded, the authors extrapolate a claim about a neural "map" used by other brain areas. Even when the examination only directly spoke to Aspect 1, the investigation of place-correlated hippocampal cells was closely connected to a view of a larger, navigational system and thus Aspect 2.

Philosophers across quite distinct intellectual traditions have championed some version of the idea that even highly complex behaviors can be understood in terms of the coordination of internal physical mechanisms. Descartes famously endeavored to rigorously specify how various behaviors are subserved by mechanisms in the nervous system (Descartes 1633/1972). Although many of Descartes' specific hypotheses have been disconfirmed and although he claimed certain "higher" operations were carried out by a non-physical mind (not subject to mechanistic analysis), nonetheless his work exemplifies the aim of explaining flexible behavioral responsiveness by appeal to neural mechanisms. Much longer ago, a broadly mechanistic view of complex behavior is also expressed in ancient India, in the Samkhya-Yoga school of thought. Samkhya divides the mind into three major components or organs, which are constituted by various kinds of internal physical changes and sensory-motor exchange with the world (See Patanjali's Yoga-Sutras, especially the Samkhya-Karika text and associated commentaries, available in Radhakrishnan and Moore 1957). There are arguably significant analogies between the Samkhya approach and contemporary western thinking about brain-based representational mechanisms (Perrett 2001). While our observational and theoretical tools have developed immensely in the modern era, historical philosophical approaches like these articulate a project of understanding cognitive capacities in terms of physically inner processes with specific causal relations to things in the outside world. It seems to us that neuroscience means for representations to be at the center of such an explanatory project, and therefore that representation implies not just correspondence, but also having a functional role in some mechanistic process that shapes behavior.

Current philosophical accounts of representation widely assume or argue that representations figure in causal relations that mediate behavior. Various philosophers in the late 20th century were concerned to account for the sort of causal connection that, in their view, must exist between a representation, its contents, and other parts of a representational system. Fodor's articulation of this issue was influential, as was his answer in terms of "asymmetric causal dependency" between representations and their contents (Fodor 1990). A number of philosophers around the same time clearly state and wrestle with the problem of accounting for inner states that are taken to cause behavior (Dretske 1988; Horgan 1989; Millikan 1984). A lively discourse emerged around the issue and philosophers continue to debate the central ideas from these views of representation. All sides of this debate support the idea that correspondence (Aspect 1) does not suffice to classify the kind of internal states of interest. Of course, neuroscientists and philosophers do not share all of the same explanatory interests, but here there seems plainly to be an overlap in the two field's ideas about information-carrying states with a special, content-bearing role in adaptive behavior. Across competing philosophers' views, the problem has been conceived partly as being to answer what justifies calling some brain activity a representation of $F$ *other than* (perhaps in addition to) the fact that the activity is found to correlate with $F$. A causal relationship to the behavior of the representation-containing system (Aspect 2) shows up as a part of many philosophers' attempts to elucidate what makes for the content of representations.

This second aspect can be specified further. Saying that neural activity $N$ represents $F$ implies not just any effect on $F$-related behavior; it implies the effect of one or more brain mechanisms that make use of $N$ as a representation of $F$. In other words, the effect of a representation should be specifiable in terms of a mechanism in which $N$ functions as a representation. This functional description of $N$ alongside other parts of the system should support predictions about how behavior will be affected when a representation is experimentally blocked or induced. This makes $N$ explanatorily relevant in a way that other activity that correlates with $F$ is not. In Soh et al.'s model of *c elegans* chemotaxis, for example, the representation of NaCl gradient by two interneurons is described within a mechanism that describes how a worm's salt-following and turning behaviors are generated. Such a mechanism incorporates a representation by showing how a state with some content (something like "salt, *that way*") figures in a complex of processes that have some observable, behavioral output. Other parts of the *c elegans* nervous system to which the two interneurons are closely related also shape the creature's following and turning behaviors, illustrating the fact that various parts of a mechanism other than the representation of $F$ also have effects on $F$-related behavior. Therefore, when investigating a mechanism hypothesized to describe how the brain might solve some problem, it is often a challenge to disentangle which neural properties exactly implement which parts of which mechanism. In creatures somewhat more complex than *C elegans*, one also faces (more) challenging questions about how a mechanism is importantly contingent on parts of the nervous system outside of wherever it is implemented.

So in short, if some neural activity correlates with $F$ and is found to have *some* effect on $F$-related behavior, $N$ still might not be a representation of $F$. It could be that $N$ represents something more general or specific than $F$, or $N$ could be just a part of a larger representation of $F$, or $N$ could be a causal relay involved in $F$-related behaviors without representing anything in particular. Many patterns of neural activity involved in $F$-related behaviors are likely to modify a behavior in some fashion, but to say that all simultaneously play the role of representing $F$ in the containing system would be to use the "representation" in a way that implies an incredible amount of redundancy, and without any explanatory relevance beyond correspondence alone. A hypothesis of a functional role for $N$ in some containing system involved in behavior is, we suggest, part of what makes $N$ merit the status of "representation" while other causally implicated neural states do not. In practice, this means that neural activity with overlapping information and causal roles must not be seen in parallel streams: if other brain regions have similar activity and a similar effect on behavior, this activity is either also implicated in the representation, or such activity is not distinctive enough to have a specific representational role.

The question of what defines a "mechanism" or "mechanistic explanation" has generated an influential and ongoing philosophical literature (Cartwright 1999; Machamer et al 2000; Hardcastle 2002, Bechtel & Abrahamson 2005; Andersen 2011; Baumgartner & Gebhardter 2016; Glennan 2017), and several contributors take special interest in mechanisms in the brain (Craver 2007; Kaplan 2011; Harbecke 2014; Boone & Piccinnini 2016). These philosophers broadly offer views of the central role of mechanism in scientific understanding in widely varying domains, and of how to think about relations between different mechanisms in a complex system. An issue of primary concern is that of understanding relations between functional descriptions at quite different spatial and temporal scales. Neuroscientists in particular are interested in explaining phenomena within an individual neuron, within a neural circuit, across brain-areas, and across animals, and many questions arise about how explanations at such different levels relate to one another. Not all mechanistic explanations involve representations, but these philosophers do not all agree about the conditions under which a mechanism does use a representation. In Soh et al.'s (2018) model of *c elegans*, for example, the NaCl gradient information is relayed in some form through much of the nervous system, but some feature of how that information is processed between two interneurons must warrant locating the representation there. Philosophers have tried in different ways to clarify the difference between mechanistic processes that merely carry information and those that perform their causal role *by* carrying information. A particular point of contention has been has been whether mechanistic and representation-involving explanations include dynamic systems models of neural systems (and if so, which ones) (Van Gelder 1995, Chemero 2001, Kaplan & Bechtel 2011, Zednik 2011, Silberstein & Chemero 2013, Beer & Williams 2015).[6] One fairly common idea is that the information carried by some neural activity must be, in some measure, decoupled from other parts of the system if it is to figure in a representation. However resolving this issue in detail is beyond our present scope. The point here is that this philosophical literature supports the idea that neuroscience means to use representations in mechanistic explanations, and therefore that representations are partly defined by their causal role in the behavior of a representation-using system.

A relatively common way of describing the relationship between representations and mechanisms in the brain does not cohere with the recognition that representation involves Aspect 2. A recent review article centers on "brain-wide representations" in a few organisms, including *c elegans*, suggesting that features of ongoing behavior are represented in much of their nervous systems (Kaplan & Zimmer 2020). The authors apply the term "representation" to all neural activity that encodes features of ongoing behavior, suggesting that whether and how a representation is used is not essential to its status as such. On this approach, far more of the *c elegans* nervous system than the two interneurons described in the research we discussed earlier would be counted as representing the worm's NaCl-tracking behavior. Similarly, some descriptions of place cells in the hippocampus suggest they represent spatial features regardless of whether they are used to interact with those spatial features. For instance, an overview by Hartley et al. (2014) seems to imply that a correspondence relation suffices for representation, but notably their paper starts by asking how an animal is able to know where it is, remember distant goals, and navigate to them. A review of the topic by Colgin (2020) highlights how many things *other than* spatial relations correspond to hippocampal activity in humans, and relates these findings in terms of hippocampal "representations" of various nonspatial aspects of experience, such as visual and temporal features. The terms of these authors also gives no indication that a representation of *F* implies a functional role for that representation. One philosopher who works on mechanistic explanation adopts this kind of view of place cell representations; in Bechtel's (2016) terminology, the hypothesis about what hippocampal cells represent should precede any consideration of what the representations might be used for. While he does envision a mechanistic role in navigation for place cells, in his terms the content of these representations - the fact that they are about spatial features - does not depend on them playing such a role. Here we are arguing for a different way of framing things. It is not disputed that much of the *c elegans* nervous system carries information about ongoing behavior, and no mechanistic hypothesis is needed to see that the hippocampus carries information about a spatial layout. This is just the first aspect. But when it is claimed that not just information but a "representation of food" or "place" resides in some neural activity, this, we argue, rides on the assumption that they can play a functional role related to food- or place-related behavior.

---

[6] Some philosophers think that the challenge of finding distinctive causal roles that internal states play in virtue of the information they carry cannot be met, and so they deny the explanatory value of representation altogether (Ramsey 2007; Egan 2020).

In our estimation, this second aspect is usually not examined by systems neuroscience experiments, although establishing causal relations between neural activity and behavior may be increasingly feasible with the creative application of new tools (Urai et al. 2021). While causality is often recognized as important to studying representations in neuroscience, in many settings it is not well established (Marinescu, Lawlor, and Kording 2018). Compared to estimating correlations, establishing a causal effect on behavior for some neural circuits is more difficult. There are many ways neuroscientists can probe causal relations -- including lesion and knockout experiments, electrophysiological stimulation (Isitan et al. 2020; Penfield and Rasmussen 1950), and optogenetics (Deisseroth et al. 2006). Nonetheless these experiments tend to be hard; performing targeted interventions on the right neurons, and only those neurons, in the right way, still is a challenge in many organisms. This makes it somewhat easier to make reasonably precise hypotheses about representations in a creature like *C elegans*, whose entire nervous system activity over the course of food-related locomotive behavior can be observed in some detail, and much of it intervened on at relatively low cost. Even when perturbations are possible, they are generally low-dimensional, making it hard to say much about the causal effects of components of a representation as opposed to the entire representation.

**Aspect 3: Teleology**

The third aspect says that a representation must serve some aim, goal or purpose ("telos" is ancient Greek for purpose). This aspect is highlighted by a consideration of cases involving mistakes or illusions, which deviate in a basic way from the more ordinary cases of successful perception and movement that we have considered so far. Such cases highlight the fallibility of representations, revealing an implied standard that actual representations either meet or fall short of. For a non-neural example, one normally says a map comes with a standard of accuracy; the fact that it is a map *of* (representing) a certain place implies a range of conditions that determine the extent to which it is a good map. Such accuracy conditions help explain, for instance, whether someone using the map can be expected to navigate successfully, or how they might be misled. The map might correspond with several different regions, but it helps explain the *mis*guided navigator's behavior only with respect to the region it represents, illustrating that accuracy conditions cannot be read off of the correspondence relation that the map bears to a particular place. Similarly, perceptual states are sometimes inaccurate, beliefs are sometimes false, and inferences are sometimes unwarranted by one's evidence. Representations are supposed to allow neuroscientists to help explain patterns of misperception and misdirected behavior, as well as the more mundane "good cases." That is, representations are also supposed to help explain how behavior can be elicited in inappropriate circumstances - a representation of *F* can help tell us why an organism behaved as if it encountered an *F* even though it did not. So fallibility is essential to representations that neuroscientists try to identify. This means that, supposing one has identified a pattern of neural activity $N$ whose correspondence with $F$ (Aspect 1) plays a functional role in a process that modifies $F$-related behavior (Aspect 2), something further is needed to say why $N$ represents $F$ (in error) when abnormal environmental conditions mean the mechanism fails to support the behavior. Some basis is needed for saying that $N$ is *supposed to* correspond with $F$, thereby establishing the possibility of misrepresentation. As we illustrate below, understanding neural representations to be embedded in a teleological framework (Aspect 3) accounts for this normative dimension of their explanatory significance in a way that comports with common ways of thinking in neuroscience.

To illustrate this in the context of the earlier *c elegans* example, we can first consider any teleological language used by the authors in describing the relevant processes. Soh et al. (2018) initially describe the movements involved in chemotaxis as strategies for survival, and more specifically as being for "approaching a favorable environment." A more specific description of the purpose of chemotaxis might be appropriate, at least if there are other favorable conditions that *c elegans* can sense and approach that do not relate to NaCl concentration. Perhaps "approaching nutrients" would be an improvement, but the example will serve in any case. We can imagine a mistake and a misrepresentation occurring in the following kind of case: part of the creature or environment outside of the proposed neural mechanism is affected so that the direction represented by the pair of interneurons is made to be *un*favorable. Suppose an extraneous electric current or an injury modifies the creature's sensory response so that the difference in the interneuron activity now leads the worm to swim 90 degrees to the right of where the highest concentration of NaCl is. If we have a clear enough view of the effect of the electrical current or injury, then the neural mechanism for chemotaxis and the included representation of the NaCl gradient helps explain the observed pattern of (mis-) behavior. This suggests that part of what one hopes to gain in discovering how the creature's nervous system represents the NaCl gradient is an understanding of how it might behaviorally mistake an unfavorable environment for a favorable one (or more specifically, a nutrition-scarce location for a nutrient-rich one). Our suggestion is that, since the content of the *c elegans* representation is meant to contribute to our understanding of failure modes like this, we rely on the notion of a purpose or goal that the functional process containing the representation serves.

Philosophers of mind going back to antiquity have been focused on analyzing the problem of erroneous mental states, which is also commonly known as the problem of "intentionality," owing to the famous use of the term by Brentano in 1874 (Caston 2019; Taieb 2018). Mental representations are supposed to help us understand how mistakes and failures of reference happen, which is why they are said to exhibit intentionality. Philosophers who hope to understand the physical nature of mind partly in terms of representations in the brain broadly see representations as intentional in this sense. Intentionality is what makes error possible, and representations are critical to the explanation of how errors occur. Thus, one is in no position to say when a physical system is representing something unless one is also in position to say when it is misrepresenting (Dretske 1986). By relying on a teleological view of the thing *for which* some neural activity and the mechanism it is part of is purposed, one can determine the difference between purpose-achieving representations and defective ones (misrepresentations).

There are two kinds of contemporary philosophical views of what determines a "telos" - often called a "proper function" of some component in a mechanistic process; one backward-looking and one forward-looking. Backward-looking, also called "etiological" approaches essentially hold that a component's proper function is what it did in the past to contribute to its current presence in the larger system. On this approach, a component's proper function explains "why it is there," in a causal, evolutionary sense, or in other words the function is what the component was naturally selected for

(Wright 1973; Millikan 1989; Papineau 1987; Dretske 1988; Godfrey-Smith 1994; Neander 1995; Garson 2017; Shea 2018). In the (organismal) systems of interest, this idea of selection often involves that of reproductively established families. For example, given a chain of self-reproducing systems that all have a heart as a component, and assuming that it was past hearts' contribution to the circulation of blood that mainly explains the presence of similar hearts today, this view says that blood-circulating is the proper function of the heart. Hearts have other interesting characteristic properties that could be called functions - for instance, they make a rhythmic sound and reflect certain wavelengths of light more than others - but these do not qualify as the heart's *proper* function, on this kind of view, because these are not what it was selected for. In analogous fashion, a backward-looking teleological view of neural representation would derive the content of such representations from the selection history of the neural activity.

A challenge for philosophers who favor backward-looking teleology (especially with respect to parts of brains) is to precisely define the relevant causal (non-intentional) notion of "selection."[7] In doing so, one must address the so-called "problem of novel contents," which means accounting for proper functions involving objects too novel to have influenced reproductive selection. For instance, parts of a human's brain might represent technologies or social structures that did not exist one or a few generations ago. Relatedly, sometimes components seem to suddenly acquire new functions, as might have happened the first time wings were used for digging or hands used for signing. Champions of a backward-looking approach have proposed ways of addressing these issues, which are beyond our scope here.[8]

The alternative is a forward-looking approach, sometimes called a Goal-Contribution view. Forward-looking teleology is based on some definition of systems that exhibit goal-directed behavior, and the proper functions of components are determined by their contributions to specific goals in such systems. That is, one starts with some way of discerning which systems have goals, at least in general terms and perhaps also in formal terms. Typically some measures of mutability and robustness are taken to be evidence for the goal-directedness of some system's behavior. In other words, one asks how various are the circumstances from which the system will proceed to the outcome (goal), and how various are the perturbations or obstacles that processes toward that outcome can accommodate. Different philosophers have articulated these notions somewhat differently, but they similarly try to account for goals as outcomes that the system's behavior *would adjust to ensure* across a wide range of counterfactual scenarios (Nagel 1977; Bigelow and Pargetter 1987; Boorse 2002; Mossio, Saborido, and Moreno 2009; Cao 2012). Whatever formal approach one takes, one must also deal with the complicating possibility of one system having conflicting goals, or goals that change over time. Again, we leave more thorough analyses of these issues for a different discussion.[9] This teleology at the system-level is what determines the proper roles of functional components; a component is supposed to do whatever contributes to the system's capacity to direct itself to the goal. This provides for a (non-etiological) notion of "why the component is there," which justifies treating other features of the component as functionally irrelevant and justifies treating some activities as malfunctional. For instance, on this view one can claim that "the heart is there to circulate blood" insofar as blood-circulation contributes to some goal-directed activity of the bodily system in which the heart functions. Accordingly, some neural activity $N$ would have the function of corresponding with some $F$ only if doing so figured in an ensemble of functional components that together generate a goal-directed activity.

---

[7] We do not take it as obvious precisely what it means for some function - say, a function of corresponding with *F* - to be naturally selected for. However we leave that topic for a different discussion, noting that providing a satisfying backward-looking account depends on having a clear way of determining what selected functions are.

[8] Briefly, some have appealed to the idea of a component having a function to acquire a function, such that acquired functions can be shaped according to novel circumstances (Millikan 1984; Garson and Papineau 2019). Garson (2017) offers an account of selection in terms of differential retention in a population, which can occur without reproductive lineages and which, he argues, applies to the selection of synapses.

[9] There are forward-looking views with minimal requirements on the complexity of goal-directedness, for which systems like a prokaryotic cell, a Watt Governor, and Ashby's homeostat (1960), would all count as goal-directed (Nagel 1977; Boorse 2002). Others defend more substantive requirements, typically including mechanisms to maintain a constantly-decaying body that contains those mechanisms (Mossio, Saborido, and Moreno 2009; Bigelow and Pargetter 1987)

Whether just one of these two approaches to teleology can ultimately resolve all the cases of interest does not matter here; perhaps a pluralistic view will prove best (Preston 1998). Both views provide a frame for analyzing the components of living things that affirms the teleology that distinguishes them from other natural phenomena (Mayr 2004).

Arguably, a teleological view operates in the background of most neuroscientific research, implicit in the researchers' choices of which external variables to investigate and sometimes made explicit. For instance, in the Soh et al. study of chemotaxis in *c elegans*, the authors assume that the relevant mechanism serves the purpose of survival, which might be tied to assumptions about its selection history (backward-looking) or tied to a view of the creature as one with the goal of surviving (forward-looking). For organisms like this, whose behavioral (and thus representational) repertoire is simple enough, one can be somewhat safe from doubt about the purpose some process serves. When there are few plausible purposes to choose from, backward- and forward-looking views will tend to agree and be supported by widely shared assumptions, and so a careful consideration of one's teleological framework is not essential. In such cases, the largely implicit or vague nature of appeals to teleology does not seriously hinder communication.

However, when it comes to the study of more complex behaviors with more contextual variability, it becomes non-obvious which goals an organism pursues and which mechanisms it relies on to do so, and also non-obvious how such mechanisms may have served the reproductive successes of its ancestors. For example, we can reconsider what is represented in the rat hippocampus, or to raise even more difficulties, human hippocampus. Looking back at the classic Burgess & O'Keefe (1996) work mentioned above, we could say that the hippocampal cells are assumed to be used for navigation. We might therefore roughly say that place cell activity represents distances and directions to locations of expected reward in a known environment. That said, one might also think the cells are used to represent spatial information in the course of exploration (before the environment is known), which would suggest a representation of "place" in a slightly different sense. Complicating things further, there is evidence that hippocampal cells help support basic cognitive abilities often exercised in nonspatial domains, such as memory consolidation, planning, and imagination (Epstein et al 2017, Eichenbaum 2017, Olafsdottir et al 2018). It seems fair to say that researchers are convinced that place cells are used by larger processes that allow for spatial navigation (perhaps among other capacities), but they do not yet share a clear view of what those larger processes are. Thus in our view, studies in complex areas like this only provisionally or incompletely point to representations, since they assume the representations figure in larger functional processes for specific purposes that remain obscure. We propose that this can and should be made more evident in the way neuroscientists describe representations of "place" in the hippocampus, for example, so that we do not lose sight of the gaps in our knowledge we hope to fill.

To bring backward-looking teleology more into the foreground in neuroscience, one approach would be to rely on data on the evolutionary development of the neural process under investigation. Some recent work in this spirit relies on knowledge about the phylogenetic tree to argue for constraints on the kinds of behaviors that neuroscience should study (Cisek 2019). Although this work does not speak directly to identifying neural representations, it aligns with the general principle of looking to teleology derived from reproductive success to refine neuroscientific research. An avenue more aligned with forward-looking teleology is to start by inferring organismal goals from macroscopic behavior (Ng et al. 2000; Körding 2007), carefully observing the kinds of errors the organism is prone to make, and look for representations that seem essential to explain the observed patterns of successful and errant behavior (Krakauer et al. 2017). Another approach that centrally features forward-looking teleology investigates the structure of artificial neural networks optimized for a particular task and compares it with the brain, the idea being that how the learning goal shapes the developed neural system is essential to the workings of both artificial and biological cases (Kriegeskorte 2015; Yamins and DiCarlo 2016; Richards et al. 2019).

Understanding the function of a given brain region, neural circuit or neuron is a common aim of neuroscientific studies - insofar as talk of "functions" entails some notion of what fulfills the goals of a species or individual, this suggests a strong commitment to some form of teleology. However, teleological ideas are often hidden or ignored, as experimental data itself can usually only speak to the question of encoding or even only to correlation. This casual approach to thinking about the purposive structure of representations offers no principled basis on which to distinguish the cases of misrepresentation, wherein the observed encoding relationship breaks down but the neurons are used as they normally are, leading to inappropriate behavior. To carefully investigate representations, observations of what neural activity corresponds with should be integrated into a well-described causal and teleological framework. Our hope is that, by creating the expectation that research aiming to uncover neural representations clarify these other aspects of representation, neuroscientists can communicate more precise hypotheses and advance the pace of learning about how neural systems help generate behavior.

**Conclusion**

For a final illustration, we can succinctly apply our three-aspect view of representation to generically describe a map as a representation of spatial relations among locations. In other contexts one might think of a map as a kind of cultural artifact whose contents are defined by social convention, but let us set aside that way of thinking about representations for present purposes. Aspect 1 is exhibited by the fact that the arrangement of parts of the map corresponds to the spatial arrangement of locations in some region; the map elements were made so as to match those spatial relations, and that the former carry information about the latter. Aspect 2 is exhibited insofar as the map is understood to play a role in navigation; identifying a map puts us in position to explain how a capable user of the map could travel among the represented locations. Aspect 3 is exhibited by the map's standard of accuracy, fitting with the purposive construal of "navigation" as involving goal locations; in terms of the map we explain how map-users *successfully* navigate, and how they might go astray due to inaccuracies in the map. We have here advanced the idea that when neuroscientists aim to identify maps and other sorts of representations in the brain, they do so in a sense that involves these three aspects. Clearer communication in the field will be facilitated if neuroscientists express their findings in a way that attends to the complexity and depth of this core idea.

More broadly, we have tried to show how neuroscientific research should take seriously the philosophical assumptions that underlie it. Philosophers are not well-positioned to make specific hypotheses about how neural systems work, but they have thought long and hard about the intricate conceptual interplay among notions of correspondence, mechanism, teleology, explanation, representation and related ideas. To improve our understanding of how complex behavior depends on interactions among neurons, we should not ignore ideas from a millenia-spanning sub-discipline full of competing analyses about the source of such behavior. This paper means to exemplify the sorely needed collaboration between neuroscience and philosophy. We hope to have shown how the investigation of neural systems should be grounded in philosophical reflection on what representation involves.

**Objections and Replies**

In communications with colleagues in developing this paper, in particular on twitter, there were questions and qualms we encountered about a few key issues repeatedly. We drew on these discussions to formulate three objections that might occur to our readers, which we respond to here in the hopes of making our view more clear and convincing.

Objection 1: "In my experience and talking to the scientists and engineers I know, 'representation' only implies Aspect 1. It is a fine term for talking about correspondence, so let us just use it that way and officially detach the term from any notions of functional role or teleology."

Reply 1A: In common, non-theoretical contexts, the word "representation" connotes semantic value, not a property of the statistics of certain events happening together. Dictionary definitions of the word and associated examples standardly involve intentional depiction or designation, as in an artistic representation of a subject, or as in the case of one person *speaking for* another person and thereby representing their thoughts or interests.

Reply 1B: As discussed above, there is strong precedent in theoretical literature for holding that representation implies more than Aspect 1. We pointed out numerous areas of philosophical literature that support the idea that Aspect 1 does not suffice (see especially Dretske 1981; Fodor 1990; Bickhard 2000; Millikan 2001; Shea 2018; Baker 2021). We also noted influential accounts that describe neural representations as figuring in functional processes that mediate behavior (Decharms and Zador, 2000; Gallistel and King 2009).

Reply 1C: Aspect 1 alone can be well-described in other terms. In particular "information" and "correlate" are already commonly used labels for this purpose. Further, many neuroscientific studies use these words in ways that suggest "represent" is to be understood as importantly different from "carry information about" or "correlate with". We noted above how Aspects 2 and 3 are often at least mentioned in connection to studies of neural representation, and noted that a survey on the use of the term in neuroscience suggests that much of the field associates the term with more than Aspect 1, and suggests that a representation's content must be meaningful to the organism (Vilarroya 2017).

Objection 2: "Representations are not always involved in behavior. For example, suppose I am idly recalling the plot of a fictional story, or imagining myself in that story; the representations involved in my mental activity do not eventuate in behavior."

Reply 2A: Having a functional role in a behavioral mechanism does not mean that a representation must immediately guide the organism in interacting with the thing represented. The point is rather about the form of explanation that representations are supposed to allow for. Plausibly, the acts of recall or imagination described in the example above *do* serve general functions - these functions might involve responding to features of one's social context or planning for novel circumstances. If we were able to carefully probe such temporally extended, subtle cognitive capacities that humans have, we might expect to find that, in fact, representing the fictional scenario does modify a range of observable behavior. For instance the representations might modify how one responds to a questionnaire, or how one will react if something similar to the story occurs in the near future. Our claim is that an internal state corresponding to some plot-element $F$ must be able to help us explain something about behavior, in principle, to warrant the status of representation.

Reply 2B: The examples in the objection involve capacities that are uniquely human, which act on long time-scales, for which we have scant evidence about the brain processes involved. To improve our understanding of what representations are, in general, we should not focus on the exceptional case of an organism representing extremely abstract entities whose relation to the organism's body is quite indirect and not well known. If we can agree about what representations are like in simpler cases, such as navigating in an ecologically relevant space, then we can try to see how that understanding might be extended to more sophisticated examples.

Objection 3: "This view of representation is too demanding. One cannot expect the general run of neuroscience experiments to substantially establish Aspects 2 and 3."

Reply 3A: As we said in the introduction, our view is adamantly not that the only valuable brain research identifies what some part of the brain represents. For all we say, it can be good to probe just one aspect and leave it to later work to integrate that into a bigger picture involving the other aspects. The point is for this to be done in an unconfusing way, which requires some care in making claims about representation. A neuroscientist may have good reasons to assume or suppose the existence of a functional role for $N$ as a representation of $F$ without having that role described in nearly as much detail as the parts of the mechanism for chemotaxis in *c elegans*. This means that one may reasonably consider some neural activity as a representation without yet knowing the mechanism it figures in, provided one is committed to the idea that some behavioral mechanism or other makes use of the representation. However it is important to be clear about these theoretical commitments, on pain of being unable to tell how different results support or conflict with one another.

Reply 3B: Using neuroscientific data to establish general claims about what the brain is doing to support some behavior or cognitive process is a very demanding sort of project. It requires taking a stance, explicitly or not, on certain philosophical questions about the explanatory significance of inner states that mean something specific in the world of the animal whose nervous system it is. Representations are a central element in this prodigious explanatory picture, so one should not expect that the conceptual apparatus attached to the term would be straightforward to express linguistically or probe experimentally. It might be nice if meaningful understanding of the neural components of intelligent capacities were easy to come by, but our reflection on the main ideas at play indicates that they are not.

**Acknowledgements:** This work was supported by NIH (1-R01-EB028162-01, R01EY021579) and by the University of Pennsylvania Office of the Vice Provost for Research.

______________________________